\documentclass[11pt]{article}
\usepackage{latexsym} 
\usepackage{fullpage}

\usepackage{amsfonts}
\def\Real{\mathbb{R}}

\def\01{\{0,1\}}
\newcommand{\ceil}[1]{\lceil{#1}\rceil}

\newcommand{\ket}[1]{|#1\rangle}
\newcommand{\bra}[1]{\langle#1|}
\newcommand{\outp}[2]{|#1\rangle\langle#2|}

\newcommand{\inp}[2]{\langle{#1}|{#2}\rangle} 

\newcommand{\id}{I}

\newtheorem{definition}{Definition}
\newtheorem{theorem}{Theorem}

\newtheorem{proposition}{Proposition}
\newtheorem{corollary}{Corollary}

\newenvironment{proof}
{\noindent {\bf Proof. }}
{{\hfill $\Box$}\\
 \smallskip}
\newcommand{\mA}{\mathcal{A}}
\newcommand{\mB}{\mathcal{B}}
\newcommand{\mV}{\mathcal{V}}
\newcommand{\mP}{\mathcal{P}}
\newcommand{\mM}{\mathcal{M}}

\title{Entanglement in Interactive Proof Systems with Binary Answers}
\author{Stephanie Wehner\thanks{Supported by EU project RESQ IST-2001-37559 and NWO Vici grant 2004-2009.}\\
        CWI, Kruislaan 413, 1098 SJ Amsterdam, the Netherlands.\\
        {\tt wehner@cwi.nl}}

\begin{document}
\maketitle
\begin{abstract}
If two classical provers share an entangled state, the resulting interactive proof
system is significantly weakened~\cite{cleve:nonlocal}. We show that for the case 
where the verifier computes the XOR of two binary answers, the resulting proof system
is in fact no more powerful than a system based on a single quantum 
prover: $\oplus\mbox{MIP}^*[2] \subseteq \mbox{QIP}(2)$. This also implies
that $\oplus\mbox{MIP}^*[2] \subseteq \mbox{EXP}$ which was previously shown using a 
different method~\cite{cleve:nonlocalTalk}. This contrasts with an interactive
proof system where the two provers do not share entanglement. In that case,
$\oplus\mbox{MIP}[2] = \mbox{NEXP}$ for certain soundness and completeness 
parameters~\cite{cleve:nonlocal}.\\[2mm]
\end{abstract}

\section{Introduction}

Interactive proof systems have received considerable 
attention~\cite{babai:ipNexp,benor:ip,cai:ip,feige:twoProverOneRound,lapidot:ip,feige:mip}
since their introduction by Babai~\cite{babai:ip} and
Goldwasser, Micali and Rackoff~\cite{goldwasser:ip} in 1985.
An interactive proof system takes the form of a protocol
of one or more rounds between two parties, a verifier and a prover. Whereas the prover 
is computationally unbounded, the verifier is limited to probabilistic polynomial time. Both 
the prover and the verifier have access to a common input string $x$. The goal of the prover is 
to convince the verifier that $x$ belongs to a pre-specified language $L$. 
The verifier's aim, on the other hand, is to determine whether the prover's claim is indeed valid. 
In each round, the verifier sends a polynomial (in $x$) size query to the prover, who returns a 
polynomial size answer. At the end of the protocol, the verifier decides to accept, and 
conclude $x \in L$, or reject based on the messages exchanged and his own private randomness.
A language has an interactive proof if there exists a verifier $V$
and a prover $P$ such that: If $x \in L$, the prover can always convince $V$ to accept.
If $x \notin L$, no strategy of the prover can convince $V$ to accept with non-negligible probability.
IP denotes the class of languages having an interactive proof system. 
Watrous~\cite{watrous:qip} first considered the notion of \emph{quantum} interactive proof systems.
Here, the prover has unbounded quantum computational power whereas the verifier is restricted
to quantum polynomial time. In addition, the two parties can now exchange quantum messages.
QIP is the class of languages having a quantum interactive proof system.
Classically, it is known that
$\mbox{IP} = \mbox{PSPACE}$~\cite{shamir:ipPspace,shen:ipPspace}.
For the quantum case, it has been shown that 
$\mbox{PSPACE} \subseteq \mbox{QIP} \subseteq \mbox{EXP}$~\cite{watrous:qip,kitaev&watrous:qip}.
If, in addition, the verifier is given polynomial size quantum advice, the resulting class
$\mbox{QIP}/\mbox{qpoly}$ contains all languages~\cite{raz:quantumPCP}.
Let $\mbox{QIP}(k)$ denote the class where the prover and verifier are restricted to exchanging $k$ messages. 
It is known that $\mbox{QIP} = \mbox{QIP}(3)$~\cite{kitaev&watrous:qip} and
$\mbox{QIP}(1) \subseteq \mbox{PP}$~\cite{vyalyi:qma,mariott:qip}. 
We refer to~\cite{mariott:qip} for an overview of the extensive work 
done on $\mbox{QIP}(1)$, also known as $\mbox{QMA}$.
Very little is known about 
$\mbox{QIP}(2)$ and its relation to either $\mbox{PP}$ or $\mbox{PSPACE}$. 

In multiple-prover interactive proof systems the verifier can interact with multiple, computationally
unbounded provers. Before the protocol starts, the provers are allowed to
agree on a joint strategy, however they can no longer communicate during the execution of the protocol. 
Let MIP denote the class of languages having a \emph{multiple}-prover interactive proof system.
In this paper, we are especially interested in two-prover interactive proof systems as introduced
by Ben-Or, Goldwasser, Kilian and Widgerson~\cite{benor:ip}.
Feige and Lov\'asz~\cite{feige:mip} have shown that a language is in NEXP if and only if 
it has a two-prover one-round proof system, i.e. MIP[2] = MIP = NEXP. Let $\oplus\mbox{MIP}[2]$ 
denote the restricted class where the verifier's output is a function of 
the XOR of two binary answers. Even for such
a system $\oplus \mbox{MIP}[2] = \mbox{NEXP}$, for certain
soundness and completeness parameters~\cite{cleve:nonlocal}. Classical multiple-prover interactive 
proof systems are thus more powerful than classical proof systems based on a single prover, assuming
$\mbox{PSPACE} \neq \mbox{NEXP}$. 
Kobayashi and Matsumoto have considered \emph{quantum} multiple-prover interactive
proof systems which form an extension of quantum single prover interactive proof systems as described
above. Let $\mbox{QMIP}$ denote the resulting class. In particular, they showed that
$\mbox{QMIP} = \mbox{NEXP}$ if the provers do \emph{not} share quantum entanglement. 
If the provers share at most polynomially many entangled qubits the resulting class
is contained in $\mbox{NEXP}$~\cite{kobayashi:mip}.

Cleve, H{\o}yer, Toner and Watrous~\cite{cleve:nonlocal} have raised the question whether 
a \emph{classical} two-prover system is weakened when the provers are allowed to share arbitrary 
entangled states as part of their strategy, but all communication remains classical. 
We write $\mbox{MIP}^*$ if the provers share entanglement.
The authors provide a number of examples which demonstrate that the soundness condition of 
a classical proof system can be compromised, i.e. the interactive proof system is 
weakened, when entanglement is used.
In their paper, it is proved that 
$\oplus\mbox{MIP}^*[2] \subseteq \mbox{NEXP}$. Later, the same authors also showed 
that $\oplus\mbox{MIP}^*[2] \subseteq \mbox{EXP}$ using 
semidefinite programming~\cite{cleve:nonlocalTalk}. Entanglement thus clearly weakens an
interactive proof system, assuming $\mbox{EXP} \neq \mbox{NEXP}$. 

Intuitively, entanglement
allows the provers to coordinate their answers, even though they cannot use it to communicate.
By measuring the shared entangled state the provers can generate correlations which they can use 
to deceive the verifier. Tsirelson~\cite{tsirel:original,tsirel:separated} has shown that even quantum mechanics
limits the strength of such correlations.
Consequently, Popescu and Roehrlich~\cite{popescu:nonlocal,popescu:nonlocal2,popescu:nonlocal3} have raised 
the question why nature imposes such limits. To this end, they constructed a 
toy-theory based on non-local boxes~\cite{popescu:nonlocal,wim:thesis}, which are 
hypothetical ``machines'' generating
correlations stronger than possible in nature. In their full generalization, non-local boxes
can give rise to any type of correlation as long as they cannot be used to signal.
van Dam has shown that sharing certain non-local 
boxes allows two remote parties to perform any distributed computation using only a single 
bit of communication~\cite{wim:thesis,wim:nonlocal}.
Preda~\cite{preda:talk} showed that sharing non-local boxes can then allow two provers
to coordinate their answers perfectly and obtained $\oplus\mbox{MIP}_{\mbox{\tiny{NL}}} = \mbox{PSPACE}$, 
where we write $\oplus\mbox{MIP}_{\mbox{\tiny{NL}}}$ to indicate that the two provers share non-local boxes.

Kitaev and Watrous~\cite{kitaev&watrous:qip} mention that it is unlikely that a single-prover 
\emph{quantum} interactive proof system can simulate multiple classical provers, because then 
from $\mbox{QIP} \subseteq \mbox{EXP}$ and $\mbox{MIP} = \mbox{NEXP}$ it follows that
$\mbox{EXP} = \mbox{NEXP}$.

\subsection{Our Contribution}

Surprisingly, it turns out that when the provers are allowed to share entanglement
it can be possible to simulate two such classical provers by one quantum prover. This indicates that
entanglement among provers truly leads to a weaker proof system. In particular, we show
that a two-prover one-round interactive proof system where the verifier computes the XOR
of two binary answers and the provers are allowed to share an arbitrary entangled state
can be simulated by a single quantum interactive proof system with two 
messages: $\oplus\mbox{MIP}^*[2] \subseteq \mbox{QIP(2)}$. Since very little is known about
$\mbox{QIP}(2)$ so far~\cite{kitaev&watrous:qip}, we hope that our result may help to shed
some light about its relation to $\mbox{PP}$ or $\mbox{PSPACE}$ in the future.
Our result also leads to a proof that $\oplus\mbox{MIP}^*[2] \subseteq \mbox{EXP}$.

\section{Preliminaries}\label{prelim}

\subsection{Quantum Computing}

We assume general familiarity with the quantum model~\cite{nielsen&chuang:qc}.
In the following, we will use $\mV$,$\mP$ and $\mM$
to denote the Hilbert spaces of the verifier, the quantum prover and the message space respectively.
$\Re(z)$ denotes the real part of a complex number $z$.

\subsection{Non-local Games}

For our proof it is necessary to introduce the notion of (non-local) games:
Let $S$, $T$, $A$ and $B$ be finite sets, and $\pi$ a probability distribution on $S \times T$.
Let $V$ be a predicate on $S \times T \times A \times B$. Then $G = G(V,\pi)$ is the following
two-person cooperative game: A pair of questions $(s,t) \in S \times T$ is chosen at random
according to the probability distribution $\pi$. Then $s$ is sent to player 1, henceforth called
Alice, and $t$ to player 2, which we will call Bob. Upon receiving $s$, Alice has to reply 
with an answer $a \in A$. Likewise, Bob has to reply to question $t$ with an answer $b \in B$.
They win if $V(s,t,a,b) = 1$ and lose otherwise. Alice and Bob may agree on any kind of strategy
beforehand, but they are no longer allowed to communicate once they have received questions $s$ and $t$.
The value $\omega(G)$ of a game $G$ is the maximum probability that Alice and Bob win the game.
We will follow the approach of Cleve et al.~\cite{cleve:nonlocal} and 
write $V(a,b|s,t)$ instead of $V(s,t,a,b)$ to emphasize
the fact that $a$ and $b$ are answers given questions $s$ and $t$. 

Here, we will be particularly interested in non-local games. Alice and Bob are allowed to
share an arbitrary entangled state $\ket{\Psi}$ to help them win the game. Let $\mA$
and $\mB$ denote the Hilbert spaces of Alice and Bob respectively.
The state 
$\ket{\Psi} \in \mA \otimes \mB$ is
part of the quantum strategy that Alice and Bob can agree on beforehand. This means that for each game, 
Alice and Bob can choose a specific $\ket{\Psi}$ to maximize their chance of success.
In addition,
Alice and Bob can agree on quantum measurements. For each $s \in S$, Alice has a projective measurement described by 
$
\{X_s^a: a \in A\}
$ on $\mA$. For each $t \in T$, Bob has a projective measurement described
by 
$
\{Y_t^b: b \in B\}
$ 
on $\mB$.
For questions $(s,t) \in S \times T$, Alice 
performs the measurement corresponding to $s$ on her part of $\ket{\Psi}$ which gives her outcome 
$a$. Likewise, Bob performs the measurement corresponding to $t$ on his part of $\ket{\Psi}$ with outcome $b$.
Both send their outcome, $a$ and $b$, back to the verifier. The probability that Alice and Bob answer $(a,b) \in A \times B$ is then given by 
$$
\bra{\Psi}X_s^a \otimes Y_t^b\ket{\Psi}.
$$
The probability that Alice and Bob win the game is given by
$$
\Pr[\mbox{Alice and Bob win}] = \sum_{s,t} \pi(s,t) \sum_{a,b} V(a,b|s,t) \bra{\Psi}X_s^a \otimes Y_t^b\ket{\Psi}.
$$
The \emph{quantum value} $\omega_q(G)$ of a game $G$ is the maximum probability over all possible 
quantum strategies that Alice and Bob win. An \emph{XOR game} is a game where the value of $V$ 
only depends on $c = a \oplus b$ and not on $a$ and $b$ independently. For XOR games we 
write $V(c|s,t)$ instead of $V(a,b|s,t)$. We will use $\tau(G)$ to denote the value of the trivial
strategy where Alice and Bob ignore their inputs and return random answers $a \in_R \01$, $b \in_R \01$ 
instead. For an XOR game, 
\begin{equation}\label{trivialStrategy}
\tau(G) = \frac{1}{2} \sum_{s,t} \pi(s,t) \sum_{c \in \01} V(c|s,t).
\end{equation}

In this paper, we will only be interested in the case that $a \in \01$ and $b \in \01$. Alice and Bob's
measurements are then described by $\{X_s^0,X_s^1\}$ for $s \in S$ and $\{Y_t^0,Y_t^1\}$ for $t \in T$ respectively. 
Note that $X_s^0 + X_s^1 = \id$
and $Y_t^0 + Y_t^0 = \id$ and thus these measurements can be
expressed in the form of observables $X_s$ and $Y_t$ with eigenvalues $\pm 1$: 
$X_s = X_s^0 - X_s^1$ and $Y_t =  Y_t^0 - Y_t^1$. 
Tsirelson~\cite{tsirel:original,tsirel:separated} has shown that 
for any $\ket{\Psi} \in \mA \otimes \mB$ 
there exists real vectors $x_s,y_t \in \Real^N$ with $N = \min(|S|,|T|)$ such 
that $\bra{\Psi}X_s \otimes Y_t\ket{\Psi} = \inp{x_s}{y_t}$. 
Conversely, if $dim(\mA) = dim(\mB) = 2^{\ceil{N/2}}$ and $\ket{\Psi} \in \mA \otimes \mB$ is a maximally entangled state,
there exist observables $X_s$ on $\mA$, $Y_t$ on $\mB$ such that
$\inp{x_s}{y_t} = \bra{\Psi}X_s \otimes Y_t\ket{\Psi}$. See~\cite[Theorem 3.5]{tsirel:hadron} for a detailed construction.

\subsection{Interactive Proof Systems}

\subsubsection{Multiple Provers}
It is well known~\cite{cleve:nonlocal,feige:mip}, that two-prover one-round interactive 
proof systems with classical communication can be modeled as (non-local) games. Here, Alice and Bob 
take the role of the two 
provers. The verifier now poses questions $s$ and $t$, and evaluates the resulting answers.
A proof system associates with each string $x$ a game $G_x$, where $\omega_q(G_x)$ determines the 
probability that the verifier accepts
(and thus concludes $x \in L$). The string $x$, and thus the nature of the game $G_x$ is known to both
the verifier and the provers. Ideally, for all $x \in L$ the value of $\omega_q(G_x)$ is close to one,
and for $x \notin L$ the value of $\omega_q(G_x)$ is close to zero. It is possible to extend the game
model for MIP[2] to use a randomized predicate for the acceptance predicate $V$. This corresponds to $V$
taking an extra input string chosen at random by the verifier. However, known applications of MIP[2] proof
systems do not require this extension~\cite{feige:stateOfArt}. Our argument in Section~\ref{result} 
can easily be extended to deal with randomized predicates. Since $V$ is not a randomized 
predicate in~\cite{cleve:nonlocal}, we here follow this approach.

In this paper, we concentrate on proof systems involving two provers, one round of communication,
and single bit answers. The provers are computationally unbounded, but limited by the laws of
quantum physics. However, the verifier is probabilistic polynomial time bounded.
As defined by Cleve et al.~\cite{cleve:nonlocal},
\begin{definition}
For $0 \leq s < c \leq 1$, let $\oplus\mbox{\emph{MIP}}_{c,s}[2]$ denote the class of all languages
$L$ recognized by a classical two-prover interactive proof system of the following form:
\begin{itemize}
\item They operate in one round, each prover sends a single bit in response to the verifier's question,
and the verifier's decision is a function of the parity of those two bits.
\item If $x \notin L$ then, whatever strategy the two provers follow, the probability that the verifier 
accepts is at
most $s$ (the \emph{soundness} probability).
\item If $x \in L$ then there exists a strategy for the provers for which the 
probability that the verifier accepts is at least $c$ (the \emph{completeness} probability).
\end{itemize}
\end{definition}

\begin{definition}
For $0 \leq s < c \leq 1$, let $\oplus\mbox{\emph{MIP}}^*_{c,s}[2]$ denote the class corresponding
to a modified version of the previous definition: all communication remains classical, but 
the provers may share prior quantum entanglement, which may depend on $x$, and perform 
quantum measurements.
\end{definition}

\subsubsection{A Single Quantum Prover}

Instead of two classical provers, we now consider a system consisting of a single quantum prover $P_q$ 
and a quantum polynomial time verifier $V_q$ as defined by Watrous~\cite{watrous:qip}. Again, the quantum 
prover $P_q$ is computationally unbounded, however, he is limited by the laws of quantum physics.
The verifier and the prover can communicate over a quantum channel. In this paper, we are only interested
in one round quantum interactive proof systems: the verifier sends a single quantum message to the prover,
who responds with a quantum answer. We here express the definition of $\mbox{QIP}(2)$~\cite{watrous:qip} in 
a form similar to the definition of $\oplus\mbox{MIP}^*$:

\begin{definition}
Let $\mbox{\emph{QIP}}(2,c,s)$ denote the class of all languages
$L$ recognized by a quantum one-prover one-round interactive proof system of the following form:
\begin{itemize}
\item If $x \notin L$ then, whatever strategy the quantum prover follows, the probability that
the quantum verifier accepts is at most $s$.
\item If $x \in L$ then there exists a strategy for the quantum prover for which the probability
that the verifier accepts is at least $c$.
\end{itemize}
\end{definition}

\section{Main Result}\label{result}

We now show that an interactive proof system where the verifier is restricted to computing the XOR of
two binary answers is in fact no more powerful than a system based on a single quantum prover.
The main idea behind our proof is to combine two classical queries into one quantum query,
and thereby simulate the classical proof system with a single quantum prover. Recall that the two provers can use an 
arbitrary entangled state as part of their strategy. 
For our proof we will make use of the following proposition shown in~\cite[Proposition 5.7]{cleve:nonlocal}:
\begin{proposition}[CHTW]\label{clevesTheorem}
Let $G(V,\pi)$ be an XOR game and let
\newline$N = min(|S|,|T|)$. Then
$$
w_q(G) - \tau(G) = \frac{1}{2}\max_{x_s,y_t}\sum_{s,t}\pi(s,t)\left(V(0|s,t) - V(1|s,t)\right)\inp{x_s}{y_t},
$$
where the maximization is taken over unit vectors 
$$
\{x_s \in \Real^N: s \in S\} \cup \{y_t \in \Real^N: t \in T\}.
$$
\end{proposition}

\begin{theorem}\label{xormipINqip}
For all $s$ and $c$ such that $0 \leq s < c \leq 1$,
$\oplus\mbox{\emph{MIP}}_{c,s}^*[2] \subseteq \mbox{\emph{QIP}}(2,c,s)$
\end{theorem}
\begin{proof}
Let $L \in \oplus\mbox{MIP}_{c,s}^*[2]$ and let $V_{e}$ be a verifier witnessing this fact.
Let $P_{e}^1$ (Alice) and $P_{e}^2$ (Bob) denote the two provers sharing entanglement.
Fix an input string $x$. As mentioned above, interactive proof systems can be modeled
as games indexed by the string $x$. It is therefore sufficient to show that there exists 
a verifier $V_{q}$ and a quantum prover $P_q$, such that $w_{sim}(G_x) = w_q(G_x)$, where
$w_{sim}(G_x)$ is the value of the simulated game. 

Let $s$,$t$ be the questions that $V_{e}$ sends to the two provers $P_{e}^1$ and $P_{e}^2$ 
in the original game. 
The new verifier $V_{q}$ now constructs the following state in $\mV \otimes \mM$
$$
\ket{\Phi_{init}} = \frac{1}{\sqrt{2}}(\underbrace{\ket{0}}_{\mV}\underbrace{\ket{0}\ket{s}}_{\mM} + 
\underbrace{\ket{1}}_{\mV}\underbrace{\ket{1}\ket{t}}_{\mM}),
$$
and sends register $\mM$ to the single quantum 
prover $P_{q}$\footnote{If questions $s$ and $t$ are always orthogonal, 
it suffices to use $\frac{1}{\sqrt{2}}(\ket{0}\ket{s} + \ket{1}\ket{t})$.} 

We first consider the honest strategy of the prover. 
Let $a$ and $b$ denote the answers of the two classical provers to questions $s$ and $t$ 
respectively. The quantum prover now transforms the state to
$$
\ket{\Phi_{honest}} = \frac{1}{\sqrt{2}}((-1)^a \underbrace{\ket{0}}_{\mV}\underbrace{\ket{0}\ket{s}}_{\mM} + 
(-1)^b \underbrace{\ket{1}}_{\mV}\underbrace{\ket{1}\ket{t}}_{\mM}),
$$
and returns register $\mM$ back to the verifier. 
The verifier $V_{q}$ now performs
a measurement on $\mV \otimes \mM$ described by the following projectors
\begin{eqnarray*}
P_0 &=& \outp{\Psi^+_{st}}{\Psi^+_{st}} \otimes I\\
P_1 &=& \outp{\Psi^-_{st}}{\Psi^-_{st}} \otimes I\\
P_{reject} &=& I - P_0 - P_1,
\end{eqnarray*}
where $\ket{\Psi^{\pm}_{st}} = (\ket{0}\ket{0}\ket{s} \pm \ket{1}\ket{1}\ket{t})/\sqrt{2}$.
If he obtains outcome ``reject'', he immediately aborts and concludes that the quantum
prover is cheating. If he obtains outcome $m \in \01$, the verifier concludes
that $c = a \oplus b = m$. Note that $\Pr[m = a \oplus b|s,t] = 
\bra{\Phi_{honest}} P_{a \oplus b} \ket{\Phi_{honest}} = 1$, so the verifier can reconstruct
the answer perfectly. 

We now consider the case of a dishonest prover.  
In order to convince the verifier, the prover applies a transformation
on $\mM \otimes \mP$ and send register $\mM$ back to the verifier. We show that for any
such transformation the value of the resulting game is at most $w_q(G_x)$: Note that the 
state of the total system in $\mV \otimes \mM \otimes \mP$ can now be described as
$$
\ket{\Phi_{dish}} = \frac{1}{\sqrt{2}}(\ket{0}\ket{\phi_s} + \ket{1}\ket{\phi_t})
$$
where $\ket{\phi_s} = \sum_{u \in S' \cup T'} \ket{u} \ket{\alpha_u^s}$ and 
$\ket{\phi_t} = \sum_{v \in S' \cup T'} \ket{v} \ket{\alpha_v^t}$ with
$S' = \{ 0s | s \in S\}$ and $T' = \{ 1t | t \in T\}$. Any transformation
employed by the prover can be described this way.
We now have that
\begin{eqnarray}\label{crude1}
\Pr[m = 0|s,t] &=& \bra{\Phi_{dish}}P_0\ket{\Phi_{dish}} = \frac{1}{4}(\inp{\alpha_s^s}{\alpha_s^s}
+ \inp{\alpha_t^t}{\alpha^t_t}) + \frac{1}{2}\Re(\inp{\alpha_s^s}{\alpha_t^t})\\\label{crude2}
\Pr[m = 1|s,t] &=& \bra{\Phi_{dish}}P_1\ket{\Phi_{dish}} = \frac{1}{4}(\inp{\alpha_s^s}{\alpha_s^s}
+ \inp{\alpha_t^t}{\alpha^t_t}) - \frac{1}{2}\Re(\inp{\alpha_s^s}{\alpha_t^t})
\end{eqnarray}
The probability that the prover wins is given by
$$
\Pr[\mbox{Prover wins}] = \sum_{s,t} \pi(s,t) \sum_{c \in \01} V(c|s,t) \Pr[m = c|s,t].
$$
The prover will try to maximize his chance of success by maximizing 
$\Pr[m = 0|s,t]$ or $\Pr[m = 1|s,t]$. We can therefore restrict ourselves to 
considering real unit vectors for which $\inp{\alpha_s^s}{\alpha_s^s} = 1$ and $\inp{\alpha_t^t}{\alpha_t^t} = 1$.
This also means that $\ket{\alpha_s^{s'}} = 0$ iff $s \neq s'$ and $\ket{\alpha_t^{t'}} = 0$ iff $t \neq t'$.
Any other strategy can lead to rejection and thus to a lower probability of success.
By substituting into Equations~\ref{crude1} and~\ref{crude2}, it follows that the probability 
that the quantum prover wins the game when he avoids rejection is then
\begin{equation}\label{prob}
\frac{1}{2} \sum_{s,t,c} \pi(s,t) V(c|s,t) (1 + (-1)^c \inp{\alpha_s^s}{\alpha_t^t}).
\end{equation}
In order to convince the verifier, the prover's goal is to choose real 
vectors $\ket{\alpha_s^s}$ and $\ket{\alpha_t^t}$ which maximize Equation~\ref{prob}.
Since in $\ket{\phi_s}$ and $\ket{\phi_t}$ we sum over $|S'| + |T'| = |S| + |T|$ 
elements respectively, the dimension of $\mP$ need not exceed $|S| + |T|$.
Thus, it is sufficient to restrict the maximization to vectors in $\Real^{|S| + |T|}$. 
In fact, since we are interested in maximizing the inner product of two vectors from the sets
$\{\alpha_s^s: s \in S\}$ and $\{\alpha_t^t: t \in T\}$,
it is sufficient to limit the maximization of vectors to $\Real^N$ with 
$N = \min(|S|,|T|)$~\cite{cleve:nonlocal}: Consider the projection of the vectors $\{\alpha_s^s: s \in S\}$
onto the span of the vectors $\{\alpha_t^t: t \in T\}$ (or vice versa). 
Given Equation~\ref{prob}, we thus have 
$$
w_{sim}(G_x) = \max_{\alpha_s^s,\alpha_t^t} \frac{1}{2} \sum_{s,t,c} \pi(s,t) 
V(c|s,t) (1 + (-1)^c \inp{\alpha_s^s}{\alpha_t^t}),
$$
where the maximization is taken over vectors $\{\alpha_s^s \in \Real^N: s \in S\}$,
and $\{\alpha_t^t \in \Real^N: t \in T\}$.
However, Proposition~\ref{clevesTheorem} and Equation~\ref{trivialStrategy} imply that
$$
w_{q}(G_x) = \max_{x_s,y_t} \frac{1}{2} \sum_{s,t,c} \pi(s,t) V(c|s,t) (1 + (-1)^c \inp{x_s}{y_t})
$$
where the maximization is taken over unit vectors $\{x_s \in \Real^N: s \in S\}$ and 
$\{y_t \in \Real^N: t \in T\}$. We thus have
$$
w_{sim}(G_x) = w_q(G_x)
$$
which completes our proof.
\end{proof}

\begin{corollary}\label{plusmipINexp}
For all $s$ and $c$ such that $0 \leq s < c \leq 1$, $\oplus\mbox{\emph{MIP}}_{c,s}^*[2] \subseteq \mbox{\emph{EXP}}$
\end{corollary}
\begin{proof}
This follows directly from Theorem~\ref{xormipINqip} and the result that 
$\mbox{QIP(2)} \subseteq \mbox{EXP}$~\cite{kitaev&watrous:qip}.
\end{proof}

\section{Discussion}

It would be interesting to show that this result also holds for a proof system where the 
verifier is not restricted to computing the XOR of both answers, but some other boolean function. 
However, it remains unclear what the exact value of a binary game would be. 
The approach based on vectors from Tsirelson's results does not work for binary games.
Whereas it is easy to construct a single quantum query which allows the verifier to compute an arbitrary 
function of the two binary answers with some advantage, it thus remains unclear how the value of the 
resulting game is related to the value of a binary game.
Furthermore, mere classical tricks trying to obtain the value of a binary function from XOR itself seem to confer
extra cheating power to the provers.  

Examples of non-local games with longer answers~\cite{cleve:nonlocal}, such as the Kochen-Specker or the 
Magic Square game, seem to make it even easier for the provers to cheat by taking advantage of 
their entangled state. Furthermore, existing proofs that $\mbox{MIP}=\mbox{NEXP}$ break down if
the provers share entanglement. It is therefore an open question whether $\mbox{MIP}^* = \mbox{NEXP}$ or,
what may be a more likely outcome, $\mbox{MIP}^* \subseteq \mbox{EXP}$. 

Non-locality experiments between two spacelike separated observers, Alice and Bob, can be cast in the form 
of non-local games. For example, the experiment based on the well known CHSH inequality~\cite{chsh:nonlocal}, 
is a non-local game with binary answers of which the verifier computes the XOR~\cite{cleve:nonlocal}. Our 
result implies that this non-local game can be simulated in superposition by a single prover/observer:
Any strategy that Alice and Bob might employ in the non-local game can be mirrored by the single prover
in the constructed ``superposition game'', and also vice versa, due to Tsirelson's 
constructions~\cite{tsirel:original,tsirel:separated} mentioned earlier. 
This means that the ``superposition game'' corresponding to the non-local CHSH game is in fact 
limited by Tsirelson's inequality~\cite{tsirel:original}, even though it itself has no non-local character.
Whereas this may be purely coincidental, it would be interesting to know its physical interpretation, if any.
Perhaps it may be interesting to ask whether Tsirelson type inequalities have any consequences on
local computations in general, beyond the scope of these very limited games. 

\section{Acknowledgments}
Many thanks go to Julia Kempe, Oded Regev and Ronald de Wolf for useful discussions.
I would also like to thank Richard Cleve for very useful comments on an 
earlier draft. Thanks 
to Daniel Preda for his talk at the CWI seminar~\cite{preda:talk} about interactive provers using 
generalized non-local correlations
which rekindled my interest in provers sharing entanglement. Many thanks also to Boris
Tsirelson for sending me a copy of~\cite{tsirel:separated} and~\cite{tsirel:original},
and to Falk Unger and Ronald de Wolf for proofreading. Finally, thanks to the anonymous
referee whose suggestions helped to improve the presentation of this paper.

\end{document}